# Attachment of colloidal nanoparticles to boron nitride nanotubes

Mirjam Volkmann,[#] Michaela Meyns,[#,&] Rostyslav Lesyuk,[#] Hauke Lehmann,[#] and Christian Klinke[#*]

[#] Institute of Physical Chemistry, University of Hamburg, Martin-Luther-King-Platz 6, 20146 Hamburg, Germany

[&] Current address: Catalonia Institute for Energy Research—IREC, Jardins de les Dones de Negre 1, 08930 Sant Adrià de Besòs, Spain

**ABSTRACT:** There is a strong interest to attach nanoparticles non-covalently to one-dimensional systems like boron nitride nanotubes to form composites. The combination of those materials might be used for catalysis, in solar cells, or for water splitting. Additionally, the fundamental aspect of charge transfer between the components can be studied in such systems. We report on the synthesis and characterization of nanocomposites based on semiconductor nanoparticles attached directly and non-covalently to boron nitride nanotubes. Boron nitride nanotubes were simply integrated into the colloidal synthesis of the corresponding nanoparticles. With PbSe, CdSe, and ZnO nanoparticles a wide range of semiconductor bandgaps from the near infrared to the ultra violet range was covered. A high surface coverage of the boron nitride nanotubes with these semiconducting nanoparticles was achieved, while it was found that a similar *in-situ* approach with metallic nanoparticles does not lead to proper attachment. In addition, possible models for the underlying attachment mechanisms of all investigated nanoparticles are presented. To emphasize the new possibilities that boron nitride nanotubes offer as a support material for semiconductor nanoparticles we investigated the fluorescence of BN-CdSe composites. In contrast to CdSe nanoparticles attached to carbon nanotubes, where the fluorescence is quenched, particles attached to boron nitride nanotubes remain fluorescent. With our versatile approaches we expand the library of BN-nanoparticle composites that present an interesting, electronically non-interacting complement to the widely applied carbon nanotube-nanoparticle composite materials.

## INTRODUCTION

Due to their extraordinary properties freestanding tubular nanomaterials have attracted an increasing interest after the discovery of carbon nanotubes (CNTs) in 1991.[1,2] Apart from CNTs, hollow cylindrical structures of BN,[3] CuS[4], $TiO_2$,[5] and $NiCl_2$,[6] have been successfully synthesized. Hexagonal BN gained attention since it is isoelectric to graphitic carbon. It also exists as three-, two-, and one-dimensional material like the carbon allotropes graphite, graphene, and CNTs. However, in contrast to the carbon allotropes, BN structures are insulators with a bandgap larger than 5.5 eV [7] and the corresponding nanomaterials are not subject to significant confinement effects. This is due to the partially ionic character of the crystal lattice. Compared to CNTs, boron nitride nanotubes (BNNTs) are interesting due to their uniform electronic bandgap that is independent of the tube chirality and its diameter. Furthermore, BNNTs exhibit a high chemical stability and resistance against oxidation and they display very high thermal conductivity and mechanical stability making them ideal nanoparticle (NP) supports for catalytic processes[8-10]. In general, NTs have been shown to be promising candidates for applications in nanoelectronics, optoelectronics, or biomedicine.[11] Especially their low dimensionality in combination with the high volume to surface ratio opens many new possibilities. Another branch of nanotechnology concerns the synthesis of inorganic NPs. Colloidal synthesis has proven to be an efficient method for the production of high-quality NPs of various sizes and shapes.[12,13] Combinations of NPs and NTs represent promising systems due to synergetic effects, such as phototransistor or solar cell.[14,15] Thus, a wide range of different strategies has been developed for the fabrication of NP-NT composites.[16]

One key approach is based on the indirect contact between the two components. Therefore, the NT sidewalls are first covalently functionalized with so-called anchor or linker groups to which the synthesized NPs couple in a second step. The disadvantage of this method is that this surface modification does not preserve the chemical structure of the surface. In an experiment following this approach, BNNTs were covalently modified with short chain linker molecules that are terminated by thiol groups, allowing for a strong binding to Au NPs.[17]

In a second more gentle strategy of indirect attachment, the BNNTs are first wrapped with suitable molecules. In contrast to the covalent modification, weak interactions such as π-π, van der Waals, or electrostatic ones play an important role in this non-covalent approach. Thereby, it prevents the disruption of the intrinsic $sp^2$-conjugation of the NT and thus, the BNNT maintains its structural integrity. Examples are Pt decorated BNNTs using PANI as wrapping agent,[18] Ag attached to biotin functionalized BNNTs,[19] or CdS NPs on BNNTs decorated with nucleotides.[20] Unfortunately, all these approaches have no direct attachment of the NPs to the NTs.

The third approach involves direct attachment of NPs onto the surface of the NTs, either through integration of the NTs into the synthesis or by attaching the produced NPs post-synthetically. Up to now, very few studies have reported on the direct sidewall functionalization of BNNTs with NPs, examples are $SnO_2$,[21,22] $Fe_3O_4$,[23] or $TiO_2$.[24]

Based on our works on CNT composites,[15,25] we examined the possibility of transferring the procedures of *in situ* attachment of NPs to BNNTs. Furthermore, we developed new methods to decorate BNNTs not only with oxide materials but also with selenide based semiconductor NPs covering a broad absorption spectrum from the NIR to the UV. For CdSe NPs we compared the effect of attachment to BN on the NP photoluminescence.

**EXPERIMENTAL SECTION**

**Materials.** Multi-wall boron nitride nanotubes (MW-BNNTs; 70%, NTL-Composites), lead oxide (PbO; ≥99.9%, Sigma-Aldrich), oleic acid (OA; 90%, Sigma-Aldrich), 1-octadecene (ODE; 90%, Sigma-Aldrich), selenium shots (Se; amorphous, 1-3 mm, 0.04-0.1 in, 99.99+%, Alfa Aesar, stored in a glovebox under nitrogen atmosphere), tri-$n$-octylphosphine (TOP; 97%, abcr, stored in a glovebox with nitrogen atmosphere), zinc acetate dihydrate (99%, Sigma–Aldrich), 2-phenyl ethanol (PhEt; ≥99.9%, Sigma-Aldrich), potassium hydroxide (KOH; ≥85%, Carl Roth), cadmium oxide (CdO; 99.99+%, abcr), octadecylphosphonic acid (ODPA; 98%, Alfa Aesar), trioctylphosphine oxide (TOPO; 98%, Merck), 1,2-dichloroethane (DCE; 99.5%, Merck), platinum acetylacetonate (Pt(acac)$_2$; 99%, abcr), hexadecanediol (HDD; 90%, Sigma–Aldrich), oleylamine (OAm; 80-90% $C_{18}$ content, Acros Organics), diphenyl ether (DPE; 99%, Sigma–Aldrich), dicobalt octacarbonyl (Co$_2$(CO)$_8$; stabilized with 1-10% of hexane, ≥90% Sigma–Aldrich), 1,2-dichlorobenzene (DCB; 99%, Acros Organics), silver acetate (99.99+%, Sigma–Aldrich), gold(III) chloride (AuCl$_3$; 99%, Sigma–Aldrich), methanol (MeOH; p.A., VWR), toluene (p.A., VWR). All chemicals were used without further purification.

**Methods.** *Synthesis of PbSe-BNNT composites.* For the reaction 45 mg (0.20 mmol) lead oxide and 128 µL (0.40 mmol) oleic acid (OA) were dissolved in 8 mL 1-octadecene (ODE). The mixture was heated to a temperature of 180 °C until an optically clear solution was obtained. Afterwards, the reaction mixture was allowed to cool down to 80 °C and stirred for 1 h under vacuum conditions. Then, 2 mL of a BNNT suspension in ODE obtained by sonication of 10 mg BNNTs in 10 mL ODE for 5 min were added and conditioned again for 1 h. Further, it was heated up to 140 °C and 0.42 mL (0.42 mmol Se) pure selenium dissolved in tri-$n$-octylphosphine (TOP) (1 M) was rapidly injected. For the growth process the temperature was kept constant at 130 °C. After 24 h the reaction was stopped by cooling down and the obtained composites were washed several times with toluene.

*Synthesis of CdSe-BNNT composites.* For the reaction 25 mg (0.19 mmol) cadmium oxide and 193 µL (0.61 mmol) OA were dissolved in 8 mL ODE. The mixture was heated to a temperature of 280 °C until an optically clear solution was obtained. Afterwards, the reaction mixture was allowed to cool down to 80 °C and stirred for 1 h under vacuum conditions. Then, 2 mL of a BNNT suspension in ODE obtained by sonication of 10 mg BNNTs in 10 mL ODE for 5 min were added and conditioned again for 1 h. Further, it was heated up to 235 °C and 0.42 mL (0.42 mmol Se) pure selenium dissolved TOP (1 M) was rapidly injected. For the growth process the temperature was kept constant at 225 °C. After 24 h the reaction was stopped by cooling down and the obtained composites were washed several times with toluene.

*Synthesis of ZnO-BNNT composites.* For the reaction 45 mg (0.20 mmol) zinc acetate dihydrate and 65 µL (0.20 mmol) OA were dissolved in 8 mL 2-phenyl ethanol (PhEt). The mixture was heated to a temperature of 115 °C until an optically clear solution was obtained. Afterwards 2 mL of a BNNT suspension in PhEt obtained by sonication of 10 mg BNNTs in 10 mL PhEt for 5 min were added. Also at this temperature 1.1 mL (0.4 M) potassium hydroxide dissolved in PhEt was rapidly injected. For growth process the temperature was kept constant at 160 °C. After 24 h the reaction was stopped by cooling down and the obtained composites were washed several times with toluene.

*Transfer synthesis of CdSe-BNNT composites.* The preparation of the CdSe-BNNT nanocomposites was carried out by introducing adaptations to a method published by Juárez *et. al.*[25] For the reaction a mixture of 25 mg (0.19 mmol) cadmium oxide, 0.14 g (0.42 mmol) octadecylphosphonic acid (ODPA) and 3.0 g trioctylphosphine oxide (TOPO) was heated to a temperature of 120 °C for 30 min to degassing the mixture. To form the Cd-ODPA-complex the mixture was heated further to 280 °C under nitrogen flow. After 1 h the developed optically clear solution was cooled down to 80 °C to inject 2 mL of a BNNT suspension in toluene. The NT suspension was prepared by sonication of 10 mg BNNTs in 10 mL toluene for 5 min. The solvent was fully removed under vacuum before 10 µL (0.13 mmol) 1,2-dichloroethane (DCE) was inject. The mixture was then heated to a temperature of 265 °C and 0.42 mL (0.42 mmol Se) of pure selenium dissolved in TOP (1 M) was rapidly injected. For the growth of the particles the temperature was lowered to 255 °C. After 24 h the reaction was stopped by cooling down and the obtained composites were washed several times with toluene.

*Transfer synthesis of ZnO-BNNT composites.* The preparation of the ZnO-BNNT nanocomposites was conducted similarly to the synthesis described by Chanaewa *et. al.*[15] For the reaction 270 mg (1.23 mmol) zinc acetate dihydrate was mixed with 8 mL methanol (MeOH) followed by the injection of 2 mL of a BNNT suspension in MeOH. The NT suspension was prepared by sonication of 10 mg BNNTs in 10 mL MeOH for 5 min. Afterwards the mixture was heated to 60 °C. At this temperature 6.5 mL (0.4 M) potassium hydroxide dissolved in MeOH was injected. A color change from colorless to milky white was observed immediately. After a reaction time of 24 h the reaction was stopped by cooling down and the obtained composite was washed several times with MeOH.

*Synthesis of Pt-BNNT composites.* Pt NPs were synthesized according to a method of Ritz *et al.*[26] with minor modifications. For the reaction 66 mg (0.17 mmol) platinum acetylacetonate, 44 mg (0.17 mmol) 1,2-hexadecanediol (HDD), (0.05 mL (0.23 mmol) oleylamine (OAM), and 2 mL (6 mmol) OA as well as 8 mL diphenyl ether (DPE) were mixed together. This mixture was heated to a temperature of 80 °C for 60 min under vacuum conditions. Afterwards, 2 mL of a BNNT suspension in DPE obtained by sonication of 10 mg BNNTs in 10 mL DPE for 5 min were added and conditioned again for 30 min. Then, the temperature was raised to 160 °C under nitrogen atmosphere and 64 mg (0.18 mmol) dicobalt octacarbonyl dissolved in 0.6 mL 1,2-dichlorobenzene (DCB) was injected under vigorous stirring. The resulting black dispersion was stirred for 24 h and then cooled to room temperature. The obtained composites were washed several times with toluene.

*Synthesis of Ag-BNNT composites.* Ag NPs were synthesized according to a method of Peng *et al.*[27] with minor modifications. For the reaction 50 mg (0.3 mmol) silver acetate were mixed with 1.1 mL (3.3 mmol) OAm in 8 mL toluene. Then, 2 mL of a BNNT suspension in toluene obtained by sonication of 10 mg BNNTs in 10 mL toluene for 5 min were added. Under nitrogen protection, the mixture was heated to 110 °C under magnetic stirring. The solution was kept at this temperature for 24 h and cooled down to room temperature. The obtained composites were washed several times with toluene.

*Synthesis of Au-BNNT composites.* Au NPs were synthesized according to a method of Shen *et al.*[28] with minor modifications. For the reaction 98 mg (0.29 mmol) $HAuCl_4 \cdot 3\,H_2O$ were mixed with 1.1 mL (3.3 mmol) OAm in 8 mL toluene. Then, 2 mL of a BNNT suspension in toluene obtained by sonication of 10 mg BNNTs in 10 mL toluene for 5 min were added. Under nitrogen protection, the mixture was heated to 65 °C under magnetic stirring. The solution was kept at this temperature for 24 h and cooled down to room temperature. The obtained composites were washed several times with toluene.

**Characterization.** Transmission electron micrographs were obtained with a JEOL JEM 1011 microscope with a thermal emitter operated at an accelerating voltage of 100 kV. Powder X-ray Diffration measurements were carried out with a Philips X'Pert PRO MPD with Bragg Brentano geometry and a Cu(K) X-ray source emitting at 0.154 nm. Optical measurements were performed with a confocal laser scanning microscope (CLSM) FV 1000 from Olympus.

## RESULTS AND DISCUSSION

**Syntheses of semiconductor NPs on BNNTs by the oleate approach.** We studied the attachment of metal chalcogenide NPs to BNNTs. Exemplarily, we investigated PbSe and CdSe as semiconductors (SC) with a small bandgap in the near IR of $E_{gap, PbSe}$ = 0.26 eV[29] and one in the visible range of $E_{gap, CdSe}$ = 1.7 eV[30] respectively, as well as the metal oxide ZnO as an example for a semiconductor absorbing in the UV range with a bandgap of $E_{gap, ZnO}$ = 3.3 eV.[31]

In general, the different composites were synthesized in one-pot reactions based on metal-oleate complexes as illustrated in Scheme 1.

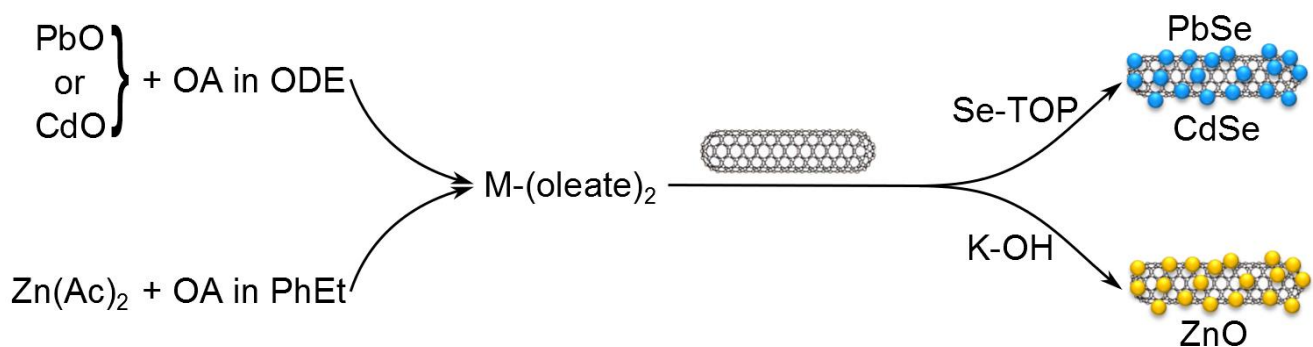

**Scheme 1.** Schematic illustration of the synthetic route of SC-NP-BNNT composites. All SC NPs were synthesized in a one-pot synthesis from previously formed metal-oleate-complexes. After the integration of the BNNTs, NP nucleation is induced by injecting Se-TOP. In case of ZnO hydrolysis of the precursor is induced by injecting KOH.

The selenides were obtained by a hot-injection reaction.[32] Briefly, BNNTs were added to PbO or CdO complexed by OA in ODE followed by the rapid injection of Se dissolved in TOP at 140 °C for PbSe or 235 °C for CdSe. The growth temperature was set 10 °C lower than the injection temperature and the reaction was terminated after 24 h. The Pb/Se/OA ratios were 1:2:2, whereas the Cd/Se/OA ratios were 1:2:3.

The Zn-(oleate)$_2$ precursor was formed by reacting zinc acetate dihydrate with OA in PhEt at a temperature of 115 °C. The Zn/OA ratios were 1:1. BNNTs were added to this solution followed by basic hydrolysis and NPs nucleation induced by injecting KOH in PhEt at a temperature of 115 °C to promote the NP formation. Soon after injection the temperature was set to 160 °C and stirred also for 24 h. In all cases the obtained composites were separated from the free NPs by several washing steps in toluene.

Figure 1 depicts representative transmission electron microscopy (TEM) images of PbSe, CdSe and ZnO attached to BNNTs after 1 h and at the end of the reaction after 24 h, respectively and the corresponding X-ray powder diffraction (XRD). Low resolution TEM images are provided in the Supporting Information, Figure S1.

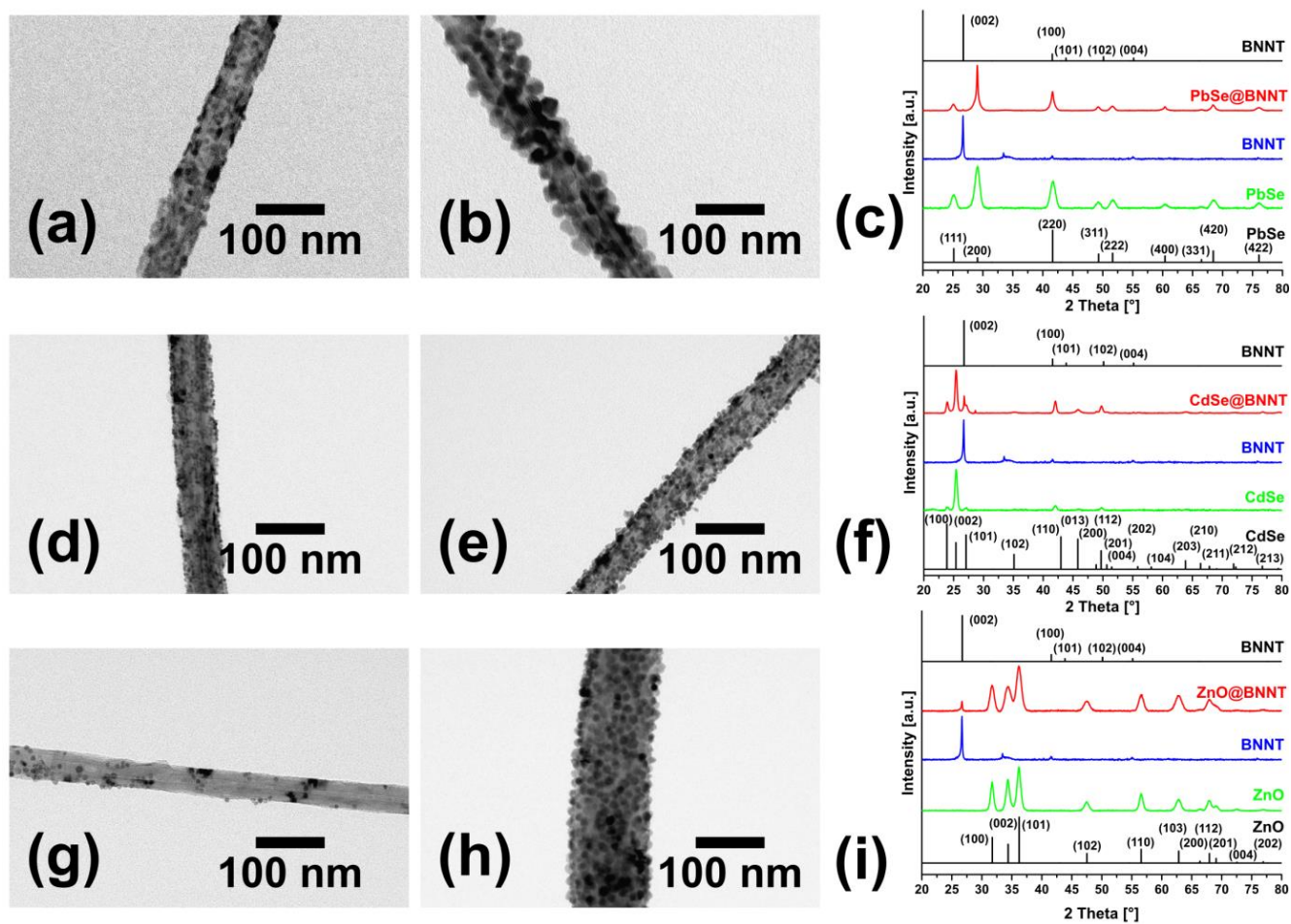

**Figure 1.** TEM images of NP-NT-composites obtained at different reaction times and the corresponding XRD. BNNTs covered with PbSe obtained after (a) 1 h and (b) 24 h (Pb:OA 1:2, 130 °C). (c) XRD patterns of pure, cubic phase PbSe NPs (green), pristine BNNTs (blue), and PbSe-BNNT-composites (red). With CdSe obtained after (d) 1 h and (e) 24 h (Cd:OA 1:4, 225 °C). (f) XRD patterns of pure hexagonal phase CdSe NPs (green), pristine BNNTs (blue), and CdSe-BNNT-composites (red). With ZnO obtained after (g) 1 h and (h) 24 h (Zn:OA 1:1 and 160 °C). (i) XRD patterns of pure hexagonal phase ZnO NPs (green), pristine BNNTs (blue), and ZnO-BNNT-composites (red).

To optimize the degree of coverage of BNNTs by NPs, various parameters were investigated. In the following, the influence of reaction time, OA concentration and temperature are discussed for the three different systems. Since PbSe and CdSe were synthesized via the same synthesis route both systems are discussed simultaneously. For ZnO it is different due to the utilization of zinc acetate and the NP formation by hydrolysis. Therefore, it will be discussed separately later on.

In the early stages (1 h) of the PbSe reaction the BNNTs are coated with a thin layer of nanostructures, as shown in Figure 1a. Both, the crystalline structure and the degree of coverage, strongly depend on the reaction time. After 24 h at a growth temperature of 130 °C and a ratio between Pb/Se/OA of 1:2:2 crystalline NPs covered fully the BNNTs. A typical TEM image of those composites is displayed in Figure 1b.

The crystalline nature of the pure BNNTs, pure PbSe NPs, and PbSe-BNNT composites after 24 h was confirmed by recording XRD. The XRD patterns are depicted in Figure 1c. All the diffraction peaks of PbSe NPs, in solution and attached, belong to the cubic structure (JCPDS-ICDD card no. 00-006-0354). It is difficult to see the characteristic BN (JCPDS-ICDD card no. 00-034-0421) peaks in the composite diffractogram because they overlap with the main peaks of PbSe. Only the (002) peak at $2\theta = 26.7$ ° is clearly visible and indicates the existence of BNNTs in the sample. Furthermore, the XRD patterns show no significant shift in the signal of the composite compared to those of pure PbSe or pure BNNTs. In comparison, Han *et al.* observed a notable shift of XRD peaks for $SnO_2$ functionalized BNNTs.[21] They suggested the formation of Sn-N bonds or electrostatic tube-particle interactions.

In order to minimize the effect of a decreasing concentration of $Pb^{2+}$ and $Se^{2-}$ (ripening) and to achieve a high coverage by NPs these two species were used in large excess compared to the amount of BNNTs. As a consequence thereof, the NPs grow not only on the BNNTs but also in solution (this is true also for the CdSe and the ZnO syntheses). Free NPs and composites can be separated by centrifugation. Remarkably, the particles which grow in solution and thus can be separated during sample cleaning have a more cubic shape (see Supporting Information Figure S2a) while the attached particles grow into pyramids during the synthesis, despite their identical cubic crystal structures. The reason for the different shape can thus be attributed to the preference of different surface facets ((100) for cubes and (111) for pyramids). This suggests two different growth mechanisms. Free NPs evolve to their thermodynamically favored cubic shape due to the rock salt crystal structure of PbSe. However, the attached NPs nucleate on the surface of the BNNTs and growth is governed by the mutual interface.

We observed slightly different growth with CdSe. At a reaction time of 1 h, the BNNTs often formed bundles wrapped with organics. This prevents the successful direct attachment of the NPs on the BNNTs. Therefore, the deposition was limited to a few distributed NPs on the BNNT walls, shown in Figure 1d. The organics wrapped around the BNNTs disappear with increasing reaction time. After a reaction time of 24 h at 225 °C and a ratio between Cd/Se/OA of 1:2:3 a uniform layer of NPs around the BNNTs is formed that finally covers the entire surface of the BNNTs, compare Figure 1e. Under these conditions the attached NPs do not grow in their preferential lattice structure. The NPs grown in solution are polydisperse and form bigger agglomerates due to ripening (see Figure S2b), which is also reflected in absorption spectroscopy (Figure S3): The excitonic shoulder of the CdSe-BNNT composites is a bit more distinct, which might be an indication for a more narrow size distribution of CdSe NCs on BNNTs.

The XRD pattern obtained for pristine BNNTs, pure CdSe NPs, and CdSe-BNNT composites, displayed in Figure 1f, again show no significant shift in the signal of the composite compared to those of the pure components. The diffractogram of CdSe (JCPDS-ICDD card no. 01-077-2307) exhibits the typical peak at $2\theta = 45.8°$, which is the (103) peak indicating that the crystals grew in a hexagonal lattice. The BNNT (JCPDS-ICDD card no. 00-034-0421) peaks around 2θ = 26.7 °, 41.5 °, 55.0 °, and 75.9 ° can be assigned to the typical (002), (100), (004), and (110) facets.

For ZnO we observed that a few pyramidally shaped NPs are attached to the outer surface of the BNNTs 1 h after nucleation, as shown in Figure 1g. Compared to Figure 1h after 24 h at a growth temperature of 160 °C and a ratio between Zn/OA of 1:1 these NPs have neither changed in size nor in shape. However, a longer reaction time increases the degree of coverage. NPs grown freely in solution are also pyramidally shaped but bigger and polydisperse in comparison to the attached NPs (see Figure S2c).

The corresponding XRD patterns for samples of pristine BNNTs, pure ZnO NPs, and ZnO-BNNT composites are displayed in Figure 1i. According to JCPDS card No. 01-79-2205 the ZnO NPs exhibit a hexagonal wurtzite structure. Only the (002) peak at 2θ = 26.7 ° is clearly visible and indicates the existence of BNNTs in the sample here, too. Also no significant shift of the peak is observed. In all three cases the crystal structure of the attached and the free nanoparticles is the same. Nevertheless the BN surface seems to possess a shape selective influence. This might be due the hexagonal structure of the BN nanotubes which harmonizes with the hexagonal (111) facet in PbSe and the (001) facets in CdSe and ZnO, respectively. Thus differences between shape and homogeneity of attached and free particles may be explained by preferential attachment of similar sized and shaped particles with facets that provide maximum interaction between the atoms of the crystal lattices.

In all cases the semiconductor NPs were synthesized by a wet chemical hot-injection processes during which they are stabilized by ligands. These ligands may shield the NPs due to surface passivation so they do not attach directly to the tube. In this context two facts might play a crucial role for a successful attachment. On the one hand, the BNNT surfaces should be free from organics as those would in general prevent attachment. Although at the beginning of the reaction, an amorphous substance is visible on the surface of the BNNTs, as shown in Figure S4a, at later stages of the reaction the surface of the BNNTs appeared to be free of organics. This is one reason for choosing such a long reaction time. On the other hand a reduced passivation of the NPs surfaces with capping ligands is preferred such that the BNNTs are able to serve as a further ligand for the NPs.[33] Indeed, when we used higher ligand concentrations the BNNTs were not fully covered by NPs. In washed samples, ultrasound treatment (ultrasound bath) cannot remove the NPs from the BNNTs. In contrast, upon addition of a large amount of ligands (OA) the NPs are removable from the NTs. This hints to strong mechanical forces which can be overcome by mild chemical treatment. This is in agreement with electrostatic interactions. Covalent bonds would not allow this kind of behavior. In addition, we observed in the PbSe and CdSe syntheses with OA as capping ligand that smaller NPs attached to a higher extend to the BNNTs than larger ones.

To investigate the amorphous substance around the NTs in more detail we performed an experiment without adding Se. After 1 h the BNNTs are separated from the solvent by washing the samples in toluene. TEM inspection showed that some areas are covered with these organic products. There seem to be strong interactions between these two materials because several washing steps have not removed the organic products from the BNNTs. To assure that the organic wrapping is not composed of the Pb-(oleate)$_2$, the BNNTs were washed and reintroduced into pure ODE before injection of the Se-precursor. However, no PbSe formation could be observed, not even after several hours. Since the reaction did not take place, the wrapping polymer seems not to contain the Pb-precursor.

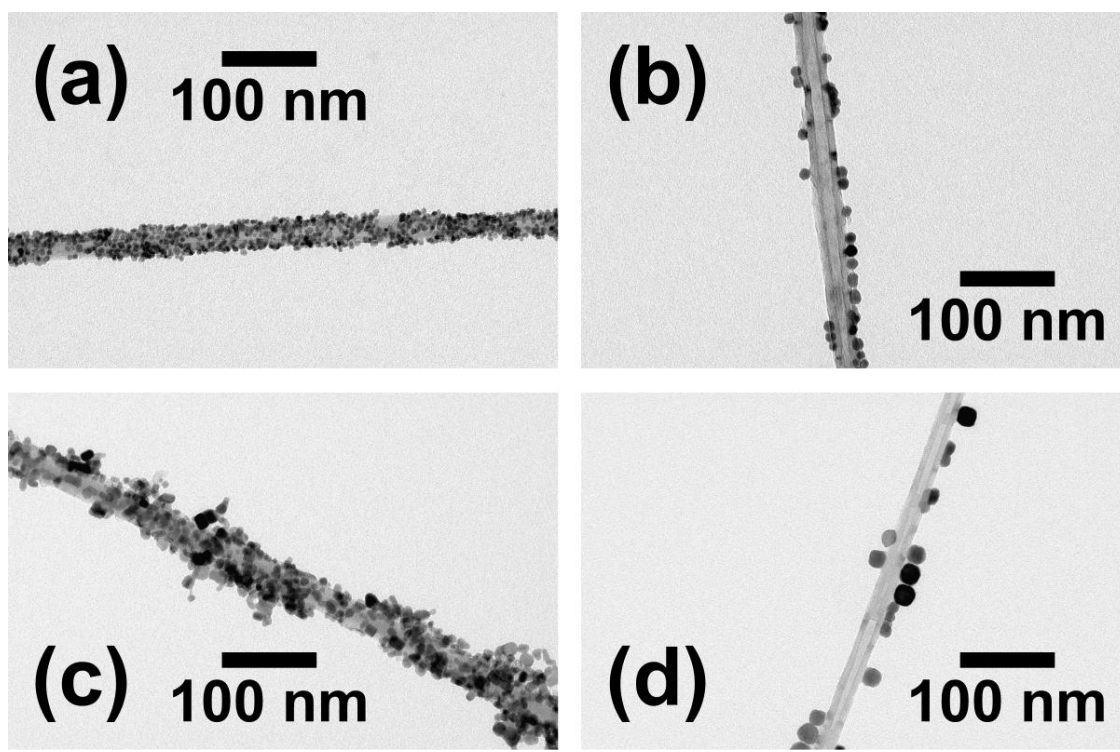

**Figure 2.** TEM images of the PbSe- and CdSe- BNNT-composite with different metal to OA ratios. PbSe (a) 1:4 and (b) 1:6. CdSe (c) 1:4 and (d) 1:6.

A higher ligand concentration reduces the activity of the monomer which results in a smaller amount of nuclei, and consequently in bigger particles.[34, 35] According to other publications, the size of the NPs strongly depends on the concentration of OA. Also in our case, if the ratio between the Pb and OA increases to 1:4, or to 1:6 the size of the obtained NPs increases as well, compare Figure 2a and b. A molar ratio of 1:2 between $Pb^{2+}$ and OA turned out to be ideal to achieve a high degree of coverage (Figure 1b). On the one hand, the amount of ligand is sufficient to form the Pb-(oleate)$_2$ complex. On the other hand, it is not too much, such that the BNNTs can act as a further stabilizer for the PbSe NPs. In the CdSe synthesis a minimum molar ratio from 1:3 between CdO and OA was necessary as otherwise the complexation to Cd-(oleate)$_2$ would not be complete as visible in form of remaining CdO powder. Again, an increased OA concentration lead to bigger NPs which showed no significant tendency to attach to the BNNTs. We investigated Cd:OA ratios of 1:4, and 1:6. Representative TEM images are depicted in Figure 2c-d.

Furthermore, when more ligands are present to stabilize the NPs the chance of the also available BNNTs to act as a further ligand for the NPs might decrease. In order to investigate this hypothesis, the composites were mixed with an excess of OA and treated in an ultrasonic bath for a couple of seconds. After the separation by centrifugation TEM images show that the majority of the NPs was removed from the BNNTs. This is not the case for an ultrasonication treatment without additional ligands.

These investigations on the ligand concentration dependency demonstrate that the attachment is favored for minimum metal to ligand ratios that keep the NP size small and that do not cover the NP surface too strongly so that it can be understood as a ligand exchange.

The influence of the reaction temperature on the formation of PbSe NPs and the attachment was studied in the range between 110 °C and 150 °C. Syntheses at a temperature of 110 °C led to inhomogeneously covered BNNTs with very small NPs of less than 2 nm in diameter which exhibit neither a uniform shape nor a uniform size as shown in Figure S5a. In comparison, a synthesis temperature of 150 °C resulted in bigger and increasingly cubic NPs, which do not show a strong tendency to attach, see Figure S5b. To achieve successful attachment, with a high degree of coverage, 130 °C turned out to be the best growth temperature.

For CdSe at growth temperatures less than 140 °C no attachment could be observed. An increase to 205 °C resulted in nearly spherical NPs which show a tendency to attach to the BNNTs, see Figure S5c. Parts of the tube sidewalls are fully covered with adsorbed NPs while other areas are only partially covered. At a reaction temperature of 245 °C the degree of coverage is remarkably reduced again while the resulting NPs grow bigger with a strong tendency to form agglomerates, compare Figure S5d. An optimal growth temperature leading to a high degree of coverage has been found to be 225 °C.

In case of ZnO NPs we utilized zinc acetate dihydrate as precursor. This leads to a different condition for growth in comparison to the metal chalcogenides syntheses. Herein the NPs are formed in a base hydrolyzed colloidal synthesis with only OA as stabilizing agent. We investigated Zn to OA ratios of 1:0.5, 1:1.5, and 1:2, illustrated in Figure 3a-c.

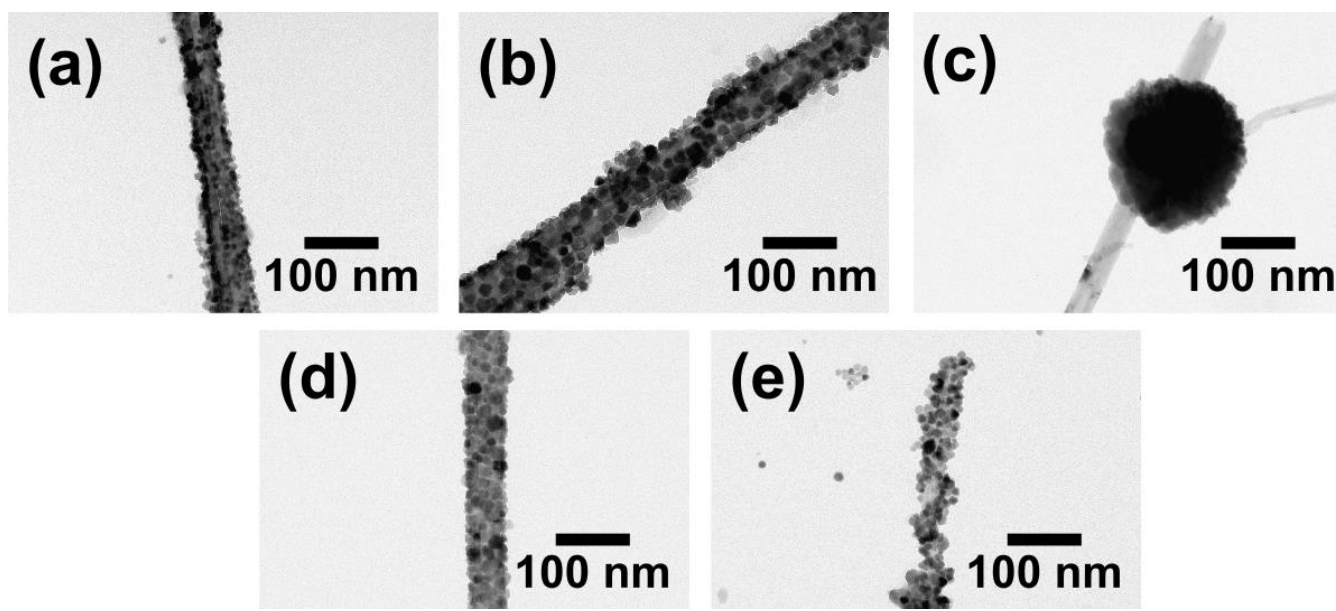

**Figure 3.** TEM images of the ZnO- BNNT-composite with different metal to OA ratios and growth temperature: (a) 1:0.5, (b) 1:1.5 and (c) 1:2 at a growth temperature of 160°C. ZnO-BNNT obtained after 24 h at growth temperatures of (d) 115 °C and (e) 200 °C.

OA is able to substitute the acetate to form Zn-(oleate)$_2$ which crucially influences the course of the reaction. Using a ratio of 1:2 the acetate can be completely substituted by oleate. As a result, a faster nucleation and growth takes place which favors large agglomerates that do not attach to the BNNTs.[36] By decreasing the amount of OA not all acetate can be replaced. In this case, we observe pyramidal NPs which are able to attach. Further decreasing the amount of OA results in smaller NPs. The highest degree of coverage is observed in case of a mixed ligand system obtained with a Zn to OA ratio of 1:1. The importance of the acetate is also confirmed by a synthesis during which the acetate was purposely removed as acidic acid by an applied vacuum. The resulting ZnO nanostructures with rod-like shape are not well attached to the BNNTs, as illustrated in Figure S4b.

Performing the ZnO synthesis at 115 °C resulted in BNNTs which are fully covered, as shown in Figure 3d. Anyhow, the attached NPs show more undefined shapes compared to Figure 1f. When the same reaction was carried out at 200 °C, the ZnO crystallization takes place very quickly in the solution. Less NPs were attached to the BNNTs while most of them stay in solution. Only after a while a higher surface coverage of the tubes with particles from the solution was achieved, illustrated in Figure 3e for a reaction time of 24 h.

These results show that it is possible to produce composites of BNNTs with attached PbSe, CdSe, or ZnO NPs by the same approach using oleate. To further investigate possibilities and limitations of NP-BNNT composite formation, we examined the transferability of the methods for the semiconducting NP-CNT composite formation developed in previous work[15,25] and the attachment of metal NPs[26] to BNNTs.

**Applying syntheses of SC NP-CNT composites to BNNT.** Regarding the fact that one material is a semiconductor (NPs) and the other one not (BNNTs), it is interesting to examine the possibility of adopting the well-established composite preparation of SC NPs with CNTs to BNNTs.

In order to compare the attachment of CdSe NPs, the previously described composite synthesis using CNTs was transferred without modifications to BNNTs.[25] Therefore, they were added to a solution consisting of CdO complexed by ODPA in TOPO followed by the injection of 10 µL DCE and the subsequent addition of the Se-TOP solution. The synthesis required 24 h at a growth temperature of 255 °C.

Furthermore, we applied the attachment method of previously investigated ZnO-CNT composites to BNNTs.[15] Therefore, the BNNTs were suspended with zinc acetate dihydrate in MeOH. A hydrolysis of the precursor was carried out at 60 °C by adding a KOH in MeOH. After 24 h of stirring the synthesis was terminated.

Representative TEM images at various times of BNNTs coated with CdSe NPs obtained by the integration of the BNNTs into the hot-injection synthesis described above is shown in Figure 4a-b.

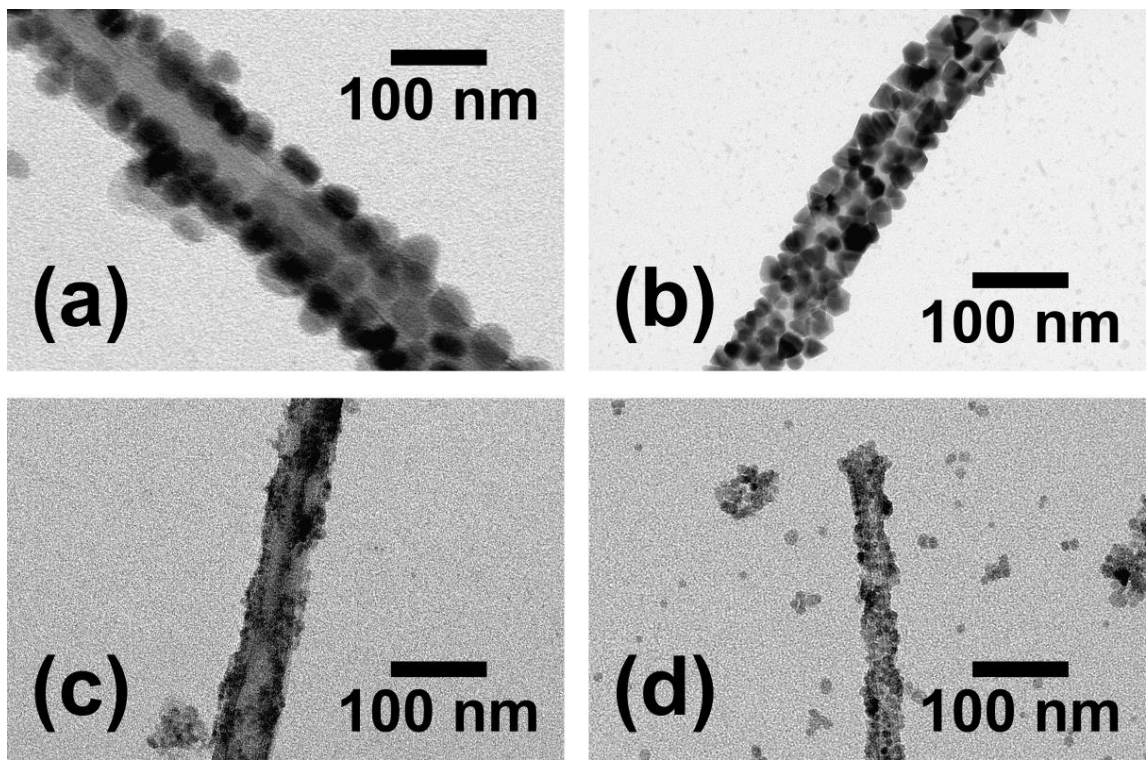

**Figure 4.** TEM images of the applying syntheses of SC NP-CNT composites to BNNT. (a) Covered with CdSe obtained after 1 h and (b) a single BNNT covered with well pyramidally shaped CdSe NPs after 24 h. BNNTs covered with ZnO obtained after (c) 1 h and (d) 24 h.

The surface of the BNNTs is fully covered with pyramidally shaped CdSe NPs. For the CNT composite synthesis method it has been reported earlier that CdSe NPs growing in solution undergo a shape transformation from rods to pyramids during the reaction in the presence of chloride and then get attached to the CNT surface.[33] This behavior is similar to our observations. In strong analogy to the CdSe-CNT composites we observe that mainly pyramidal shaped particles are capable to attach to the BNNTs. In comparison to the free NPs in the solution, those which are attached to BNNTs exhibit a more uniform size and shape. In a recent study, the role of chlorine during the reaction was investigated in more detail.[37] It turned out, that chlorine not only does influence the transformation of the particles from rods to pyramids but also plays a key role for the attachment. Chlorine ions as an atomic X-type ligand partially replace the original long chained organic ligands thereby allowing the tight attachment of especially the (000-1) facet of CdSe NPs to the CNTs.

Furthermore, it was possible to transfer our previous results on ZnO-CNT composites to BNNTs.[15] Figure 4c-d shows TEM images of ZnO BNNT composites at various stages of the growth. Already 1 h after the beginning of the hydrolysis a ZnO layer was formed around the BNNTs. With progressing reaction the NPs evolve. This suggests that the nucleation takes place directly on the walls of the BNNTs. It has also been found that particles grown freely in the solution lack uniformity and look more rod-like compared to the spherical ones grown on the surface of the BNNTs. This behavior is identical to the CNT composite synthesis. Hence, the kind of employed NTs does not seem to make any difference.

**Syntheses of noble metal NPs on BNNTs.** Again, the composite preparation approach was the integration of BNNTs into the respective colloidal syntheses. In order to ensure equal conditions, all NPs were synthesized by well-documented methods for OAm capped metal NPs.[26-28]

NPs made of noble metals such as Pt, Ag, or Au are known to show enhanced catalytic activities. An attachment to a support material with a large surface like BNNTs could be of great advantage. Additionally, BNNTs possess a high robustness against oxidation and heating. Inspired by achievements in the field of metal CNT composites[26, 38] we tried to transfer this to BNNTs where only a few studies have been reported.[17-19] All of these have in common that they are based on covalent functionalization of the BNNT walls with linker molecules. In contrast, we follow a non-covalent approach of attachment. Figure 5 depicts representative TEM images of the synthesized composites with spherical and small NPs. Compared to the SC NP investigations the degree of coverage is rather low.

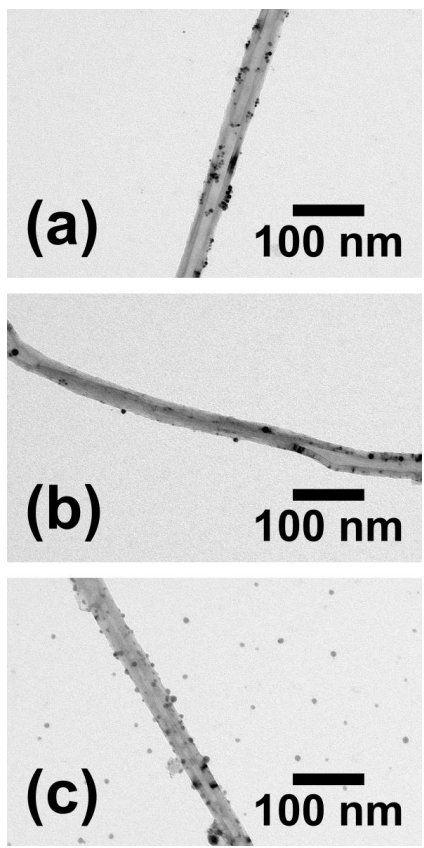

**Figure 5.** TEM images of BNNTs obtained after integrating them into the respective synthesis for (a) Pt, (b) Ag, and (c) Au NPs. In all case the composites exhibit a low degree of coverage.

This might be due to the rather small absolute binding energies between the tubes and the metal NPs, as only the nitrogen atoms of the BNNTs could bind to them.[39] It was found that the NPs are attached to the BNNTs only by chance or by *van-der-Waals* forces. Variation of the synthesis parameters such as time, amount of ligand, or temperature over a broad range does not improve the coverage. For the attachment of alloyed metallic NPs to CNTs it was found that a charge transfer is responsible for the strong adhesion between the two parts.[26] BNNTs are insulators with a large bandgap (the exact levels of the conduction and the valence band are still under debate; even negative electron affinities are discussed). Thus, it is improbable that a charge transfer from band-edge to band-edge takes place.

**Confocal microscopy investigation.** Exemplarily, we investigate the fluorescence behavior of the CdSe-OA NPs attached to BNNTs in comparison to CdSe-OA NPs attached to multiwall CNTs by confocal microscopy. Figure 6a shows an optical micrograph image of the two different NT composites. From the confocal micrograph in Figure 6b it is clearly observed that only CdSe NPs on BNNTs show strong luminescence.

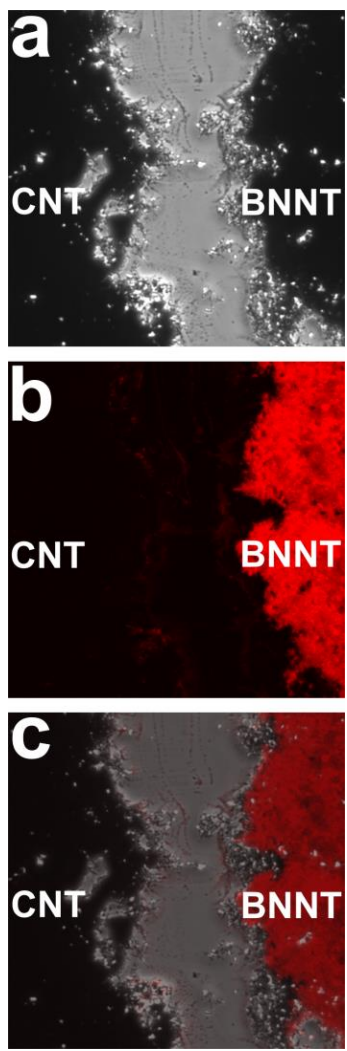

**Figure 6.** Assembled CdSe-CNT-composites (left) and CdSe-BNNT-Composites (right). (a) Optical microscopy image showing both composites. (b) Confocal fluorescent microscopy image of the same region and (c) overlay of the transmission with the fluorescent image.

The reason for the difference in the quenching behavior between CNTs and BNNTs must lie in the difference of their electronic structure. In other words, it is based on the higher charge delocalisation in the sp² lattice of carbon compared to boron nitride where the difference in electronegativity and the filling of atomic orbitals lead to electron localization preferably on nitrogen atoms. If excitons are generated in the SC NP at atomic distance from the carbon

-14-

lattice, charge transfer and its transport towards the surrounding medium can occur, thus separating electrons and holes before they are able to recombine and photoluminescence is quenched. Similar charge transport cannot occur along the insulating boron nitride structure with the result that charge recombination and thus photoluminescence occurs.

**CONCLUSIONS**

To conclude, BNNTs have been decorated successfully with semiconducting NPs of various band gaps in a non-covalent fashion. Based on similar components but slightly different conditions NP-BNNT composites can be produced by adding NTs to metal-oleate based colloidal syntheses. Such an *in-situ* synthesis is an effective approach to attach both, the metal chalcogenides PbSe and CdSe as well as the metal oxide ZnO to the BNNT surfaces. Critical factors for a high coverage of BNNTs were the ligand to metal salt ratio, the reaction temperature and time. Longer reaction times were necessary to attach NPs to the NT surface. Higher ligand contents lead to increased NP sizes that prevent effective attachment. Temperature deviations cause higher polydispersity which in turn reduces the coverage.

Well-established synthesis methods for the attachment of CdSe pyramids or ZnO NPs to CNTs have been applied successfully to the BNNT attachment. On the other hand, BN substrates behave differently to those of carbon when it comes to the attachment of metal NPs (Pt, Ag, Au) to BNNTs, where a much lower tendency to composite formation was observed. We assume that the attachment of the NPs to BNNTs is a non-covalent ligand-NP-interaction, providing an additional stabilization for the NPs. This was confirmed by ultrasonication experiments in which the addition of ligands to the purified sample combined with sonication resulted in de-attachment while composites without additional ligands remained intact.

Via fluorescence investigations, exemplarily performed on comparably produced composites of MW-CNTs and BNNTs with CdSe NPs, we found that the fluorescence of the NPs is quenched when attached to CNTs while attached to BNNTs it is conserved. This indicates that charge transfer occurs in composites with CNT while BN solely acts as a substrate or stabilizer for the NPs. With these properties the two types of composites are complementary in terms of substrate-NPs interaction that can now be chosen depending on the application and desired interaction. Based on our methods it will be possible to create composites with a large variety of materials and eventually substitute the substrate materials for other allotropes such as two-dimensional flakes. These materials may find application in various energy conversion and storage applications, especially catalysis and optoelectronics.

## ASSOCIATED CONTENT

This material is available free of charge via the Internet at http://pubs.acs.org.

Additional TEM micrographs and absorption spectroscopy.

## AUTHOR INFORMATION

**Corresponding Author**

*klinke@chemie.uni-hamburg.de

**Present Addresses**

& Catalonia Institute for Energy Research—IREC, Jardins de les Dones e Negre 1, 08930 Sant Adrià de Besòs, Spain

**Notes**

The authors declare no completing financial interest.

## ACKNOWLEDGMENT

Financial support of the European Research Council via the ERC Starting Grant "2D-SYNETRA" (Seventh Framework Program FP7, Project: 304980) as well as via the Heisenberg scholarship KL 1453/9-2 of the Deutsche Forschungsgemeinschaft (DFG) is gratefully acknowledged. The authors thank Alina Chanaewa for fruitful discussions.